\documentclass[onecolumn,english,pra]{revtex4}

\usepackage{graphicx}
\usepackage{epsfig,bm}
\usepackage{amsfonts}
\usepackage{color}
\usepackage{amsmath}
\usepackage{bigints}
\usepackage{mathrsfs}
\usepackage{url}
\usepackage{bbm}
\usepackage[colorlinks=true,allcolors=blue]{hyperref}
\usepackage{ulem}

\usepackage{ae,aecompl}
\usepackage[T1]{fontenc}
\usepackage[latin9]{inputenc}
\usepackage{color}
\usepackage{babel}
\usepackage{amsmath}
\usepackage{amssymb}
\usepackage{graphicx}
\usepackage{esint}

\definecolor{taha}{rgb}{1,0,0}

\begin{document}
\title{Q-deformed $SU(1,1)$ and $SU(2)$ Squeezed and Intelligent States and Quantum Entanglement
}
\author{Mohamed Taha \surname{Rouabah}$^{1}$, Mohamed Farouk Ghiti$^{1}$, Nouredinne Mebarki$^{1}$}
\affiliation{$^{1}$Laboratoire de Physique Math\'ematique et Physique Subatomique, Mentouri University Constantine 1, Route Ain El Bey, 25017 Constantine, Algeria\\}

\begin{abstract}
The intelligent states (IS) associated with the $su_q(1,1$) and $su_q(2)$ $q$-deformed Lie
algebra are investigated. The eigenvalue problem is also discussed.
\end{abstract}
\pacs{ 42.50.Dv, 03.65.Fd, 02.20.Hj, 02.20.Uw.}
\keywords{Intelligent states, Squeezed states. Uncertainty relation. Lie groups. Lie algebra,
Quantum groups, Entanglement.}

\maketitle

\section{Introduction}
Intelligent states are quantum states which minimize uncertainty relations for non-commuting quantum observables  \cite{Jackiw1968,Aragone1974,Aragone1976,Berghe1978,Wodkiewicz1985}. In the last years there exists a great
interest in various properties, applications and generalizations of intelligent states \cite{Wodkiewicz1985, Nieto1993, Hillery1993, Yu1994, Trifonov1994,Agarwal1994, Gerry1995}. One of the reasons for this interest is the close relationship between intelligent
states and squeezing. A generalization of squeezed states for an arbitrary dynamical
symmetry group leads to the intelligent states for the group generators \cite{Nieto1993, Hillery1993, Yu1994, Trifonov1994}. In
particular, the concept of squeezing can be naturally extended to the intelligent states
associated with the $SU(2)$ and $SU(1,1)$ Lie groups. An important possible application
of squeezing properties of the $SU(2)$ and $SU(1,1)$ intelligent states is the reduction of
the quantum noise in spectroscopy \cite{Agarwal1994} and interferometry \cite{ Hillery1993, Yu1994, Trifonov1994,Agarwal1994, Brif1994, Gerry1995} and hence improve
measurement precision. On the other hand, quantum groups are a generalization of
symmetry groups which have been used successfully in physics. A general feature of
spaces carrying a quantum group structure is that they are noncommutative and inherit
a well-defined mathematical structure from quantum group symmetries.\\
In this paper we consider a $q$-deformation of $su (1,1)$ and $su (2)$ Lie algebras and
their IS using the Dyson realization \cite{Dyson1956}.

\section{$SU_Q(1,1)$ INTELLIGENT STATES}
The $su(1,1)$ Lie algebra is spanned by the three generators $K_1, K_2$ and $K_3$ which
satisfy the following commutation relations:
\begin{equation}\label{comut:k}
[K_0, K_{\pm}] = \pm K_{\pm},	~~~~~~~ [K_+,K_-] -2K_0.
\end{equation}
The Casimir operator for any irreducible representation is $K^2 = k (k-1) I$. Thus a
representation of $su(1,1)$ is determined by the parameter $k$ called the Bergman index.
The corresponding Hilbert space is spanned by the complete orthonormal basis $\mid n,k\rangle$.
Since $SU(1,1)$ is a non compact group, all irreducible representation are of infinite
dimensions. Here we shall only deal with the representation known as the positive
discrete series in which:
\begin{eqnarray}\label{eigenvalues:K}
K_0\mid n,k \rangle &=& (n+k)\mid n,k \rangle,\nonumber \\
 K_+\mid n,k \rangle &=&\sqrt{ (n+1) (n+2k)}\mid n+1,k \rangle,\\
K_-\mid n,k \rangle &=& \sqrt{n(n + 2k-1)}\mid n-1,k \rangle\nonumber.
\end{eqnarray}

On the other hand, the uncertainty relation limits the precise knowledge of
conjugate physical quantities of a system. The state which minimize the uncertainty
relation can describe the quantum system as precisely as possible. First for a given two
self-adjoint operators $A$ and $B$, one can obtain, using the Cauchy-Schwartz inequality,
the uncertainty relation:
\begin{equation}\label{CSeneq}
\langle (\Delta A)^2\rangle \langle (\Delta B)^2\rangle = \frac{1}{4} \mid \langle [A,B] \rangle \mid ^2.
\end{equation}
where the variance and expectation value are given by $\Delta A= \langle A \rangle^2 - \langle A^2 \rangle$ and $\langle \psi \mid A \mid \psi \rangle$ respectively.
A state is called intelligent if it satisfy the strict inequality in
\eqref{CSeneq}. It is well known \cite{Aragone1974} that such state, or IS, must satisfy the eigenvalue equation
\begin{equation}\label{IS:AB}
(A + i\lambda B)\mid \psi \rangle = \eta \mid \psi \rangle,
\end{equation}
where $\lambda$ is a positive real parameter and $\eta$ a complex number.The $SU(1,1)$ intelligent states (IS) can be derived by considering the special case of Eq. \eqref{IS:AB}, where $\mid \psi \rangle$ are solutions of the eigenvalues problem
\begin{equation}\label{IS:K}
(K_1 - i\lambda K_2) \mid \psi \rangle = \eta \mid \psi \rangle.
\end{equation}
It is convenient to rewrite equation \eqref{IS:K} in term of $K_\pm$ as
\begin{equation}\label{eigenvalues:K+}
(\alpha K_+ + \beta K_-) \mid \psi \rangle = 2 \eta \mid \psi \rangle,
\end{equation}
where $\alpha = 1+ \lambda$ and $\beta = 1- \lambda$.\\

Let us expend the state $\psi$ on the basis $\mid n,k \rangle$
\begin{equation}\label{expansionpsi}
\mid \psi \rangle = \sum_{n=0}^\infty c_n(k) \mid n,k \rangle,
\end{equation}
and apply \eqref{eigenvalues:K+} to obtain the recurrence relation among the coefficients $c_n$ as fellow  \cite{AbdAl-Kader2008}
\begin{equation}
\alpha \sqrt{ (n+1) (n+2k)} c_{n+1} + \beta \sqrt{n(n + 2k-1)}c_{n-1} = 2\eta c_n.
\end{equation}\label{recurrence:1}
The $q$-deformed algebra $su_q(1,1)$ is given as \cite{Jimbo1989}
\begin{equation}
\left [ Q_0, Q_{\pm}\right ] = Q_{\pm}, ~~~~~~~ \left [ Q_+ , Q_- \right] = -2\left [Q_0 \right]_q,
\end{equation}
where the $q$-deformation is defined as
\begin{equation}\label{qDeformation}
\left [x\right ]_q = \dfrac{q^x - q^{-x}}{q-q^{-1}}.
\end{equation}
One can obtain the explicit form of the $q$-deformed generators following \cite{Curtright1990} and \cite{Oh1997} in Dyson realization as
\begin{eqnarray}\label{qDeformedOp}
Q_0 &=& K_0 = N+k,\nonumber\\
Q_- &=& K_-\sqrt{\frac{\left [N \right ]_q}{N}}, \\
Q_+ &=& K_+\sqrt{\frac{\left [N \right ]_q}{N}} \dfrac{\left [ N+2k-1\right ]_q}{N+2k-1},\nonumber
\end{eqnarray}
where N is the number operator and k is assumed to be a positive integer or half odd
integer. We require the conjugate relation $Q_-^\dag = Q_+, Q_0^\dag = Q_0$ to be independent of the realizations. These generators act on the ket as
\begin{eqnarray}\label{eigenvalues:Q}
Q_-\mid n,k \rangle &=& \sqrt{[n]_q (n+2k-1)} \mid n-1,k \rangle,\nonumber\\
Q_+ \mid n,k \rangle &=& \sqrt{[n+1]_q (n+2k)} \dfrac{[n+2k]_q}{n+2k} \mid n+1,k \rangle.
\end{eqnarray}

The $q$-deformed $SU(2)$ IS are solution of the following eigenvalue equation
\begin{equation}\label{IS:Q}
\left(\alpha Q_+ + \beta Q_-\right) \mid \psi \rangle_q = 2\eta \mid \psi \rangle_q.
\end{equation}
Substituting Eq.\eqref{expansionpsi} in Eq.\eqref{IS:Q}, and applying \eqref{eigenvalues:Q} we obtain the recurrence relation among the coefficients $c_n$ as follows:
\begin{equation}
\alpha \sqrt{[n+1]_q (n+2k)}c_{n+1} + \beta \sqrt{[n]_q (n+2k-1)}\dfrac{[n+2k-1]_q}{n+2k-1}c_{n-1} = 2\eta ~c_n.
\end{equation}\label{recurrence:2}
Assuming $c_n = (\beta_q/\alpha_q)^{\frac{n}{2}} d_n$ where $\alpha_q = \alpha \sqrt{\dfrac{[n+1]_q}{n+1}}$ and $\beta_q = \beta \sqrt{\dfrac{[n]_q}{n}}\dfrac{[n+2k-1]_q}{n+2k-1}$, we obtain:
\begin{equation}\label{polynom}
\frac{1}{2} \sqrt{(n+1)(n+2k)}d_{n+1} + \frac{1}{2} \sqrt{n(n+2k-1)}d_{n-1} - zd_n = 0,
\end{equation}
where $z^{(q)} = \dfrac{\eta}{\sqrt{\alpha_q \beta_q}}$.\\

Comparing Eq.\eqref{polynom} with the Pollaczek polynomials $P_n$ \cite{Erdelyi1953, Erdelyi1953-2, Nagel1995, Hellery1987} the solution to the eigenvalue equation (13) is directly the Pollaczek polynomials, namely,
\begin{equation}
d_n = P_n(z^{(q)},k) = i^n \left( \dfrac{\Gamma[n+2k]}{n! \Gamma[2k]} \right)^{n/2} {}_2F_1\left(-n, k+iz^{(q)}; 2k;2\right).
\end{equation}
We thus obtain the final result for the IS in the form
\begin{equation}\label{ISsuq(1,1)}
\mid \psi\rangle_q = c_0 \sum_{n=0}^\infty i^n \left( \dfrac{(1-\lambda)\Gamma[n+2k]}{n!(1+\lambda) \Gamma[2k]} \right)^{n/2} {}_2F_1\left(-n, k+iz^{(q)}; 2k;2\right) \mid n,k \rangle,
\end{equation}
where the normalization factor $c_0$ has the form
\begin{equation}\label{c0}
\mid c_0 \mid^{-2} =\sum_{n=0}^\infty i^n \left( \dfrac{(1-\lambda)\Gamma[n+2k]}{n!(1+\lambda) \Gamma[2k]} \right)^{n/2} \mid{}_2F_1\left(-n, k+iz^{(q)}; 2k;2\right) \mid^2.
\end{equation}

\section{$SU_Q (2)$ INTELLIGENT STATES}
su(2) generators satisfy the algebra
\begin{equation}
 [J_0,J_\pm] = J_\pm, ~~~~~ [J_+, J_-]= 2J_0,
\end{equation}
and the Casimir operator is given as $C = J_0(J_0 + 1) + J_- J_+$.We introduce the eigenstates
of the angular momentum as $\mid j, n-j\rangle$ where
\begin{eqnarray}\label{eigenvalues:J}
J_0\mid j, n-j\rangle &=& (n-j)\mid j, n-j\rangle,\nonumber\\
J_+ \mid j, n-j\rangle &=& \sqrt{j(j+1) -(n-j)(n-j+1)} \mid j, n-j+1\rangle,\\
J_- \mid j, n-j\rangle &=& \sqrt{j(j+1) -(n-j)(n-j-1)} \mid j, n-j-1\rangle,\nonumber
\end{eqnarray}
where $n = 0, 1, 2,\cdots, 2j$.The Hilbert space is finite dimensional with dimension $2j + 1$.
IS corresponding to $SU(2)$ generators satisfy the following eigenvalue equation \cite{Aragone1976}
\begin{equation}\label{IS:su(2)}
(J_1-i\lambda J_2) \mid \psi \rangle = \eta \mid \psi \rangle,
\end{equation}
using the raising and lowering angular momentum operators we may write Eq.\eqref{IS:su(2)} in the
following form
\begin{equation}\label{IS:J+}
(\alpha J_- + \beta J_+) \mid \psi \rangle = 2\eta \mid \psi \rangle.
\end{equation}
We may also expend $\mid \psi \rangle$ in terms of the angular momentum eigenstates $\mid j,n-1\rangle$  as
\begin{equation}\label{expansionpsi-2}
\mid \psi \rangle = \sum_{n=0}^\infty c_n(j) \mid j,n-j\rangle.
\end{equation}
Substituting Eq.\ref{expansionpsi-2} in Eq.\ref{IS:J+}, we obtain the recurrence relation for the coefficients $c_n$ .\\
The $su_q(2)$ IS can be studied in close analogy with the previous section and,
therefore, we will describe briefly $SU_q(2)$ algebra only. The $q$-deformed $su(2)$ algebra
is given as \cite{Aragone1974}
\begin{equation}\label{comut:Q-2}
\left [ Q_0, Q_{\pm}\right ] = Q_{\pm}, ~~~~~~~ \left [ Q_+ , Q_- \right] = 2\left [Q_0 \right]_q.
\end{equation}
One can obtain the explicit form of the $q$-deformed generators following \cite{Aragone1976} and \cite{Berghe1978} in
Dyson realization as
\begin{eqnarray}\label{qDeformedOp-2}
Q_0 &=& J_0 = j-N,\nonumber\\
Q_- &=& J_-\sqrt{\frac{\left [N \right ]_q}{N}}, \\
Q_+ &=& J_+\sqrt{\frac{\left [N \right ]_q}{N}} \dfrac{\left [ J-N+1\right ]_q}{J-N+1},\nonumber
\end{eqnarray}
where N is the number operator. We require the conjugate relation $Q_-^\dag = Q_+$ and $Q_0^\dag=Q_0$ to be independent of the realizations. These generators act on the ket as
\begin{eqnarray}\label{eigenvalues:Q-2}
Q_-\mid j,n-j \rangle &=& \sqrt{[n]_q (2j-n+1)} \mid j,n-j-1 \rangle,\nonumber\\
Q_+ \mid j,n-j \rangle &=& \sqrt{[n+1]_q (2j-n)} \dfrac{[2j-n]_q}{2j-n} \mid j,n-j+1 \rangle.
\end{eqnarray}
The q-deformed$SU(2)$ IS are solution of the following eigenvalue equation
\begin{equation}\label{IS:Q-2}
\left(\alpha Q_- + \beta Q_+\right) \mid \psi \rangle_q = 2\eta \mid \psi \rangle_q.
\end{equation}
Let us now consider the eigenvalue problem \eqref{IS:Q-2} we apply Eq.\eqref{eigenvalues:Q-2} to obtain the recurrence relation among the coefficients $c_n$ as follows: Substituting Eq.\eqref{expansionpsi-2} in \eqref{IS:Q-2} and
applying Eq.\eqref{eigenvalues:Q-2} we obtain the following recurrence relation among the coefficients $c_n$:
\begin{equation}
\alpha \sqrt{[n+1]_q(2j-n)} c_{n+1} + \beta \sqrt{[n]_q (2j-n+1)} \dfrac{[2j-n+1]_q}{2j-n+1} c_{n-1}
= 2\eta ~c_n.
\end{equation}
Assuming $c_n = (\beta\rq{}_q/\alpha\rq{}_q)^{\frac{n}{2}} d_n$ where $\alpha\rq{}_q = \alpha \sqrt{\dfrac{[n+1]_q}{n+1}}$ and $\beta\rq{}_q = \beta \sqrt{\dfrac{[n]_q}{n}}{\dfrac{2j-n+1}{[2j-n+1]_q}}$, we obtain:
\begin{equation}\label{polynom-2}
\frac{1}{2} \sqrt{(n+1)(n+2j)}d_{n+1} + \frac{1}{2} \sqrt{n(n+2j+1)}d_{n-1} - z\rq{}d_n = 0,
\end{equation}
where $z\rq{}^{(q)} = \dfrac{\eta}{\sqrt{\alpha\rq{}_q \beta\rq{}_q}}$.\\

Comparing Eq.\eqref{polynom-2} with the Pollaczek polynomials $P_n (\theta,b)$
\cite{Wodkiewicz1985,Nieto1993,Hillery1993}, the solution to the eigenvalue equation (29) is directly the Pollaczek polynomials, namely
\begin{equation}
d_n = P_n(z\rq{}^{(q)},j) = i^n \left( \dfrac{\Gamma[n+2j]}{n! \Gamma[2j]} \right)^{n/2} {}_2F_1\left(-n, j+iz\rq{}^{(q)}; 2j;2\right).
\end{equation}
Thus, we obtain the final result for the IS in the form
\begin{equation}\label{IS:suq(2)}
\mid \psi\rangle_q = c_0 \sum_{n=0}^{2j}  i^n \left( \dfrac{(1-\lambda)\Gamma[n+2j]}{n!(1+\lambda) \Gamma[2j]} \right)^{n/2} {}_2F_1\left(-n, j+iz\rq{}^{(q)}; 2j;2\right) \mid j,n-j+1 \rangle,
\end{equation}
where the normalization factor $c_0$ has the form
\begin{equation}\label{c0}
\mid c_0 \mid^{-2} =\sum_{n=0}^{2j}  i^n\left( \dfrac{(1-\lambda)\Gamma[n+2j]}{n!(1+\lambda) \Gamma[2j]} \right)^{n/2} \mid{}_2F_1\left(-n, j+iz\rq{}^{(q)}; 2j;2\right) \mid^2.
\end{equation}

\section{Conclusion}
A remarkable property in a $q$-deformed bipartite composite system is the existence of
a natural entangled structure for a non-classical value of the $q$-deformation parameter
($q \neq 1$).
It appears to be useful for extending the horizon of studies on entangled non
orthogonal states so as to incorporate systems with quantum algebraic symmetries
\cite{Jimbo1989}. Composite systems with quantum symmetries, such as anyons for instance, are
natural candidates.\\

\section{Acknowledgment}
We are very grateful to the Algerian Ministry of education and research, DGRSDT
and ANDRU for the financial support.

\bibliographystyle{apsrev}
\bibliography{BiblioQdeformedSIS}

\end{document}